\documentclass[cits]{PoS} 
\title{A puzzle of the $\pi^0\,\gamma$ transition
form factor---resolved?}\ShortTitle{A puzzle of the $\pi^0\gamma$
transition form factor---resolved?}
\author{Irina Balakireva\\D.~V.~Skobeltsyn Institute of Nuclear
Physics, Moscow State University, 119991, Moscow, Russia\\E-mail:
\email{balakireva.ira@gmail.com}}
\author{Wolfgang Lucha\\Institute for High Energy Physics, Austrian
Academy of Sciences, Nikolsdorfergasse 18, A-1050 Vienna,
Austria\\E-mail: \email{Wolfgang.Lucha@oeaw.ac.at}}
\author{\speaker{Dmitri Melikhov}\\Institute for High Energy
Physics, Austrian Academy of Sciences, Nikolsdorfergasse 18,
A-1050 Vienna, Austria,\\Faculty of Physics, University of Vienna,
Boltzmanngasse 5, A-1090 Vienna, Austria, and\\D.~V.~Skobeltsyn
Institute of Nuclear Physics, Moscow State University, 119991,
Moscow, Russia\\E-mail: \email{dmitri\_melikhov@gmx.de}}

\abstract{By means of QCD sum rules in the limit of ``local
duality,'' we analyze the behaviour of the form factors
$F_{P\gamma}(Q^2)$ parametrizing the amplitudes for the
transitions $\gamma\,\gamma^*\to P$ of a real photon $\gamma$ and
a virtual photon $\gamma^*$ to a pseudoscalar meson
$P=\pi^0,\eta,\eta',\eta_c$ as functions of the involved spacelike
momentum transfer $Q^2\ge0.$ Except for the findings of the {\sc
BaBar} collaboration for the $\pi^0\,\gamma$ form factor, the
experimental data for all these form factors are compatible with
saturation for large $Q^2,$ as predicted by pQCD factorization.
For the light pseudoscalar mesons $P=\pi^0,\eta,\eta',$ saturation
is observed already at relatively small
$Q^2\ge10$--$15\;\mbox{GeV}^2,$ whereas for the $\eta_c$ meson it
sets in only at larger $Q^2\ge200$--$300\;\mbox{GeV}^2.$ A recent
measurement of the $\pi^0\,\gamma$ transition form factor by the
Belle collaboration seems to resolve this disturbing puzzle as its
outcome is compatible with saturation for
$Q^2\ge10$--$15\;\mbox{GeV}^2$ and with the large-$Q^2$ behaviour
of the $\eta\,\gamma$ and $\eta'\,\gamma$ transition~form
factors.}

\FullConference{Xth Quark Confinement and the Hadron
Spectrum\\8--12 October 2012\\TUM Campus Garching, Munich,
Germany}

\begin{document}
\section{Introduction}Already a long time ago, it has been realized
that the ``two-photon fusion'' processes $\gamma^*\,\gamma^*\to P$
to some pseudoscalar meson $P=\pi^0,\eta,\eta',\eta_c$ constitute
rather crucial tests for our understanding of quantum
chromodynamics (QCD) and of the internal structure of hadrons.
Over the years, several experiments have collected impressive
amounts of information on these transition processes
\cite{cello-cleo,babar,babar2010,babar1,belle}.

As far as the theoretical description of such kind of transition
of two---in general, off-shell---photons $\gamma^*,$ with
associated polarization four-vectors $\varepsilon_{1,2},$ to a
pseudoscalar meson $P$ is concerned, the corresponding amplitude
turns out to be parametrizable by just a single form factor
$F_{P\gamma\gamma}(q_1^2,q_2^2)$:$$\langle\gamma^*(q_1)\gamma^*(q_2)|P(p)\rangle={\rm
i}\epsilon_{\varepsilon_1\varepsilon_2q_1q_2}F_{P\gamma\gamma}(q_1^2,q_2^2).$$
QCD factorization of short and long distances provides a robust
prediction for the behaviour of this form factor at asymptotically
large spacelike momentum transfers $q_1^2\equiv-Q_1^2\le0,$
$q_2^2\equiv-Q_2^2\le0$~\cite{brodsky}:
$$F_{P\gamma\gamma}(Q_1^2,Q_2^2)\to12e_c^2f_P\int_0^1\frac{{\rm
d}\xi\,\xi(1-\xi)}{Q_1^2\xi+Q_2^2(1-\xi)}.$$For convenience, we
henceforth prefer the notation $Q^2\equiv Q_2^2$ and
$0\le\beta\equiv Q_1^2/Q_2^2\le1$ (that is, $Q_2^2$ is the larger
virtuality). For the kinematics of experimental interest,
$Q_1^2\approx0$ and $Q_2^2\equiv Q^2,$ for instance, the
$\gamma\,\gamma^*\to\pi$ transition form factor
$F_{\pi\gamma}(Q^2)$ asymptotically behaves like
$Q^2F_{\pi\gamma}(Q^2)\to\sqrt{2}f_\pi,$ where
$f_\pi=130\;\mbox{MeV}$ is the charged-pion decay constant.
Similar relations arise for both $\eta$ and $\eta'$ mesons.

\section{Dispersive QCD sum rule for the form factors of the
generic transitions $\gamma^*\,\gamma^*\to P$}The analysis of the
transition $\gamma^*\,\gamma^*\to P$ within the framework of the
QCD sum-rule approach conveniently starts from the transition of
two virtual photons $\gamma^*$ to the vacuum, induced by the~quark
axial-vector current $j_{\mu}^5;$ its amplitude is found by
factorizing off the photon polarization
vectors~$\varepsilon_{1,2}$:\begin{equation}\label{AVV}
\langle0|j_{\mu}^5|\gamma^*(q_1)\gamma^*(q_2)\rangle
=e^2T_{\mu\alpha\beta}(p|q_1,q_2)\,\varepsilon^\alpha_1
\varepsilon^\beta_2,\qquad p\equiv q_1+q_2.\end{equation}We are
interested in this amplitude for $-q_1^2\equiv Q_1^2\ge0$ and
$-q_2^2\equiv Q_2^2\ge0.$ The general decomposition of
$T_{\mu\alpha\beta}$ contains \emph{four\/} independent Lorentz
structures \cite{blm2011,lm2011} but in our present study only one
enters:$$T_{\mu\alpha\beta}(p|q_1,q_2)=p_\mu\epsilon_{\alpha\beta
q_1q_2}{\rm i}F(p^2,Q_1^2,Q_2^2)+\cdots.$$For the related
invariant amplitude $F(p^2,Q_1^2,Q_2^2),$ a spectral
representation in $p^2$ for fixed $Q_1^2$ and $Q_2^2$ values may
be given in terms of its physical spectral density
$\Delta(s,Q_1^2,Q_2^2)$ and physical threshold~$s_{\rm th}$:
$$F(p^2,Q_1^2,Q_2^2)=\frac{1}{\pi}\int_{s_{\rm th}}^\infty
\frac{{\rm d}s}{s-p^2}\,\Delta(s,Q_1^2,Q_2^2).$$Perturbative QCD
(pQCD) expresses this spectral density as power series in the
strong coupling~$\alpha_{\rm s}$:$$\Delta_{\rm
pQCD}(s,Q_1^2,Q_2^2|m)=\Delta^{(0)}_{\rm pQCD}(s,Q_1^2,Q_2^2|m)+
\frac{\alpha_{\rm s}}{\pi}\Delta^{(1)}_{\rm pQCD}(s,Q_1^2,Q_2^2|m)
+\frac{\alpha_{\rm s}^2}{\pi^2}\Delta^{(2)}_{\rm
pQCD}(s,Q_1^2,Q_2^2|m)+\cdots,$$where $m$ is the mass of the quark
that propagates in that quark loop to which the two photons
couple. The well-known lowest-order term $\Delta^{(0)}_{\rm pQCD}$
arises from the graph of this one-loop quark triangle with one
axial and two vector currents at its vertices \cite{1loop}. The
two-loop $O(\alpha_{\rm s})$ correction $\Delta^{(1)}_{\rm pQCD}$
proves to vanish \cite{2loop}. The three-loop $O(\alpha_{\rm
s}^2)$ correction $\Delta^{(2)}_{\rm pQCD}$ was found to yield a
nonzero~contribution~\cite{3loop}.

In the region of small $s,$ the physical spectral density bears no
resemblance to $\Delta_{\rm pQCD}(s,Q_1^2,Q_2^2)$ as it must model
both meson pole and hadron continuum. For instance, in the $I=1$
channel one~has$$\Delta(s,Q_1^2,Q_2^2)=\pi\delta(s-m_\pi^2)\,
\sqrt{2}f_\pi\,F_{\pi\gamma\gamma}(Q^2_1,Q_2^2)+\theta(s-s_{\rm
th})\,\Delta^{(I=1)}_{\rm cont}(s,Q^2_1,Q_2^2).$$QCD sum rules
provide a possibility to relate the properties of hadronic ground
states to the spectral densities of QCD correlators. Applying this
approach in the conventional way devised by Shifman, Vainshtein,
and Zakharov proceeds along a standard routine \cite{lms1,lms2}
involving the following steps:\begin{enumerate}\item Evaluate
$F(p^2,Q_1^2,Q_2^2)$ at QCD and hadron level and equate the two
resulting representations.\item In order to suppress effects of
the hadron continuum, perform a Borel transformation $p^2\to\tau.$
\item Consider the arising sum rule in the limit of local duality
(LD), realized if the Borel parameter $\tau$ vanishes \cite{ld},
to wipe out unwanted nonperturbative power corrections increasing
with~$Q^2.$\item Implement quark--hadron duality by the customary
cut of the spectral integral at low energies.\end{enumerate}With
decay constant $f_P,$ this yields for the transition form factor a
sum rule of innocent appearance:$$\pi
f_PF_{P\gamma\gamma}(Q_1^2,Q_2^2)=\int_{4m^2}^{s_{\rm
eff}(Q_1^2,Q_2^2)}{\rm d}s\,\Delta_{\rm pQCD}(s,Q_1^2,Q_2^2|m).$$
Therein, all nonperturbative QCD phenomena are encoded in an
effective threshold $s_{\rm eff}(Q_1^2,Q_2^2);$ the actual
challenge is to design a convincing algorithm for fixing this
threshold---a nontrivial~task~\cite{lms1}.\begin{itemize}\item For
asymptotically large $Q_2^2\equiv Q^2\to\infty$ but fixed ratio
$\beta\equiv Q_1^2/Q_2^2,$ $s_{\rm eff}(Q^2,\beta)$ may be
inferred from pQCD factorization: Generally, for nonzero quark
mass $m\ne0,$ $s_{\rm eff}(Q^2\to\infty,\beta)$ depends on
$\beta;$ for $m=0,$ the factorization formula is recovered for any
$\beta$ if $s_{\rm eff}(Q^2\to\infty,\beta)=4\pi^2f_\pi^2.$\item
The na\"ive LD \emph{model\/} for the transition form factor
\emph{assumes\/} that, also for finite $Q^2,$ $s_{\rm
eff}(Q^2,\beta)$ may be sufficiently well approximated by its
asymptotic limit: $s_{\rm eff}(Q^2,\beta)=s_{\rm
eff}(Q^2\to\infty,\beta).$\end{itemize}In the LD limit, the form
factor $F_{P\gamma}(Q^2)\equiv F_{P\gamma\gamma}(0,Q^2)$ for the
transition of a pseudoscalar meson $P$ to a real ($Q_1^2=0$) and a
virtual ($Q_2^2\ne0$) photon reads, for a single massless ($m=0$)
quark flavour,\begin{equation}\label{srfp}
F_{P\gamma}(Q^2)=\frac{1}{2\pi^2f_P}\frac{s_{\rm eff}(Q^2)}{s_{\rm
eff}(Q^2)+Q^2}.\end{equation}$F_{P\gamma}(Q^2=0)$ is related to
the axial anomaly \cite{blm2011} irrespective of the behaviour of
$s_{\rm eff}(Q^2)$ near $Q^2=0.$

\section{Form factor for the transition
$\gamma^*\,\gamma^*\to\eta_c$}
\begin{figure}[t]
\includegraphics[scale=.4949]{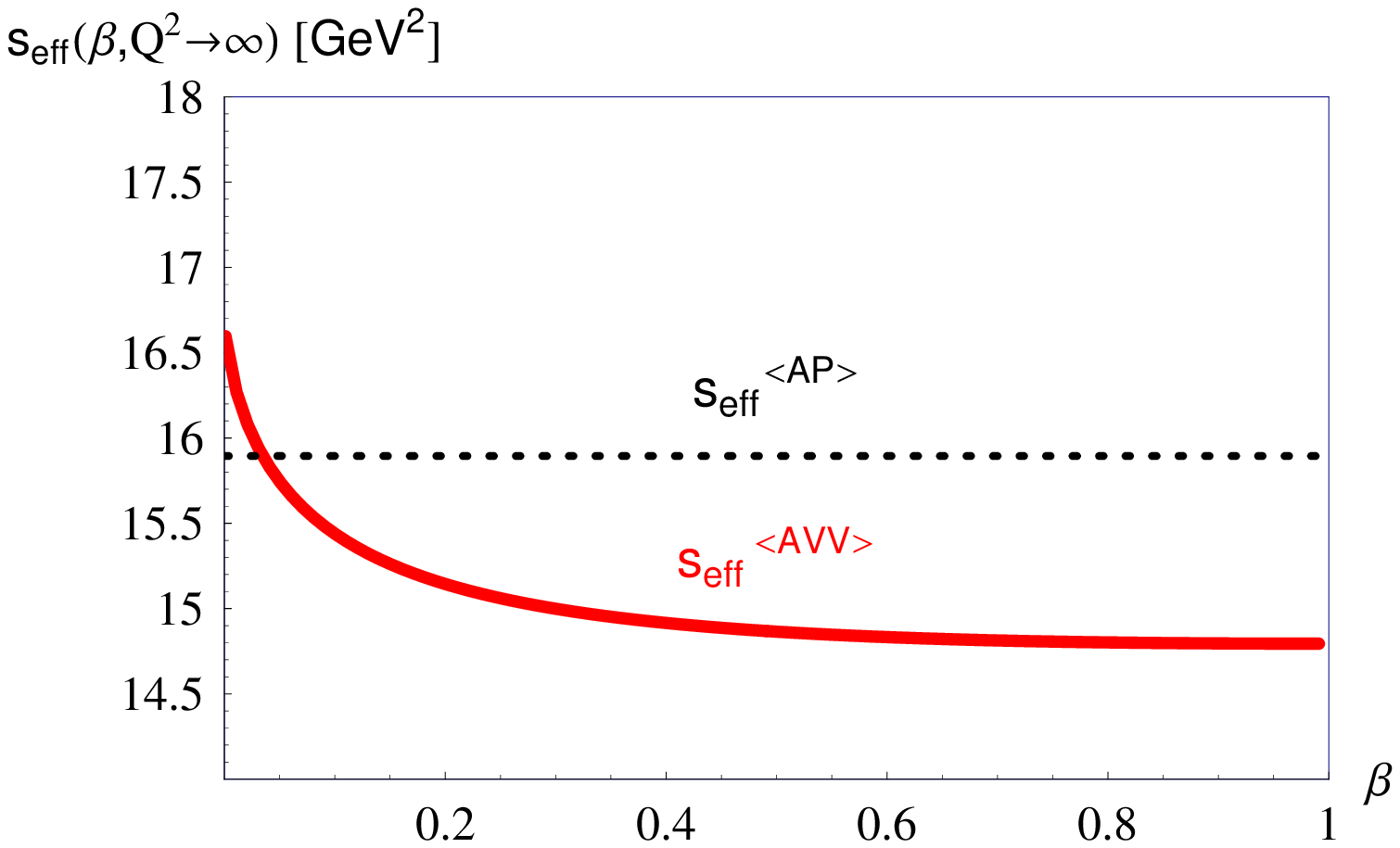}\hfill
\includegraphics[scale=.4699]{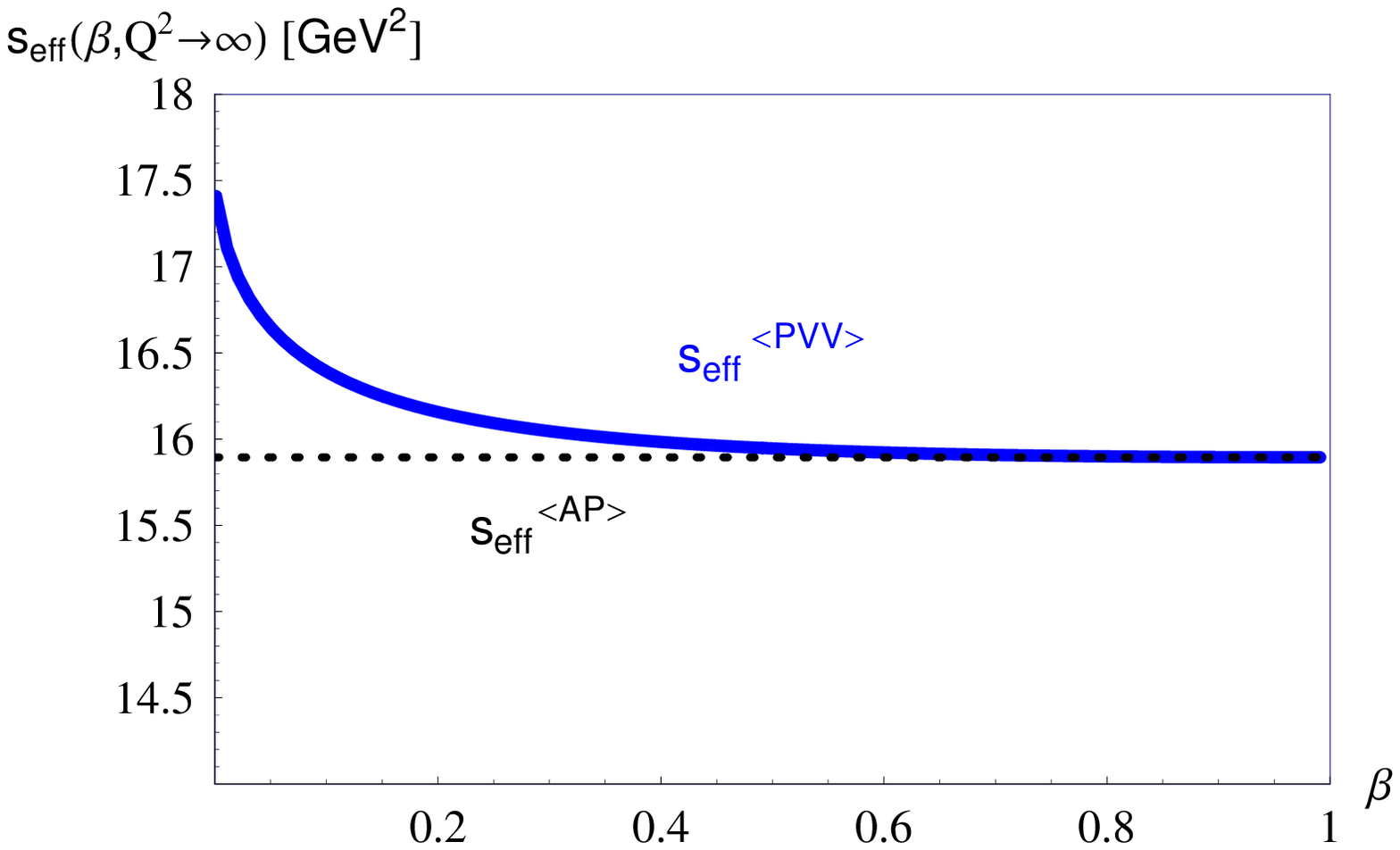}\\
\includegraphics[scale=.4935]{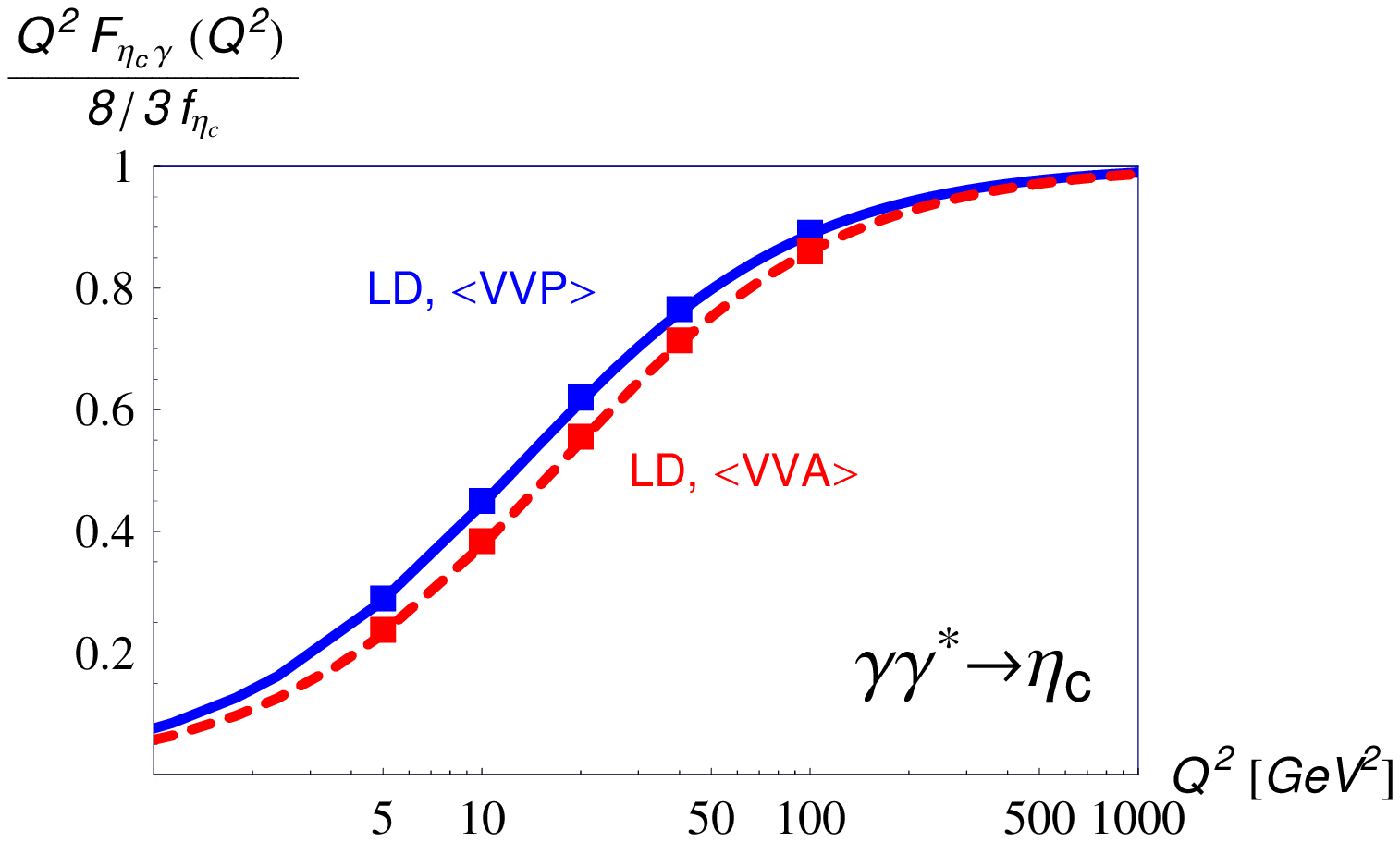}\hfill
\includegraphics[scale=.4998]{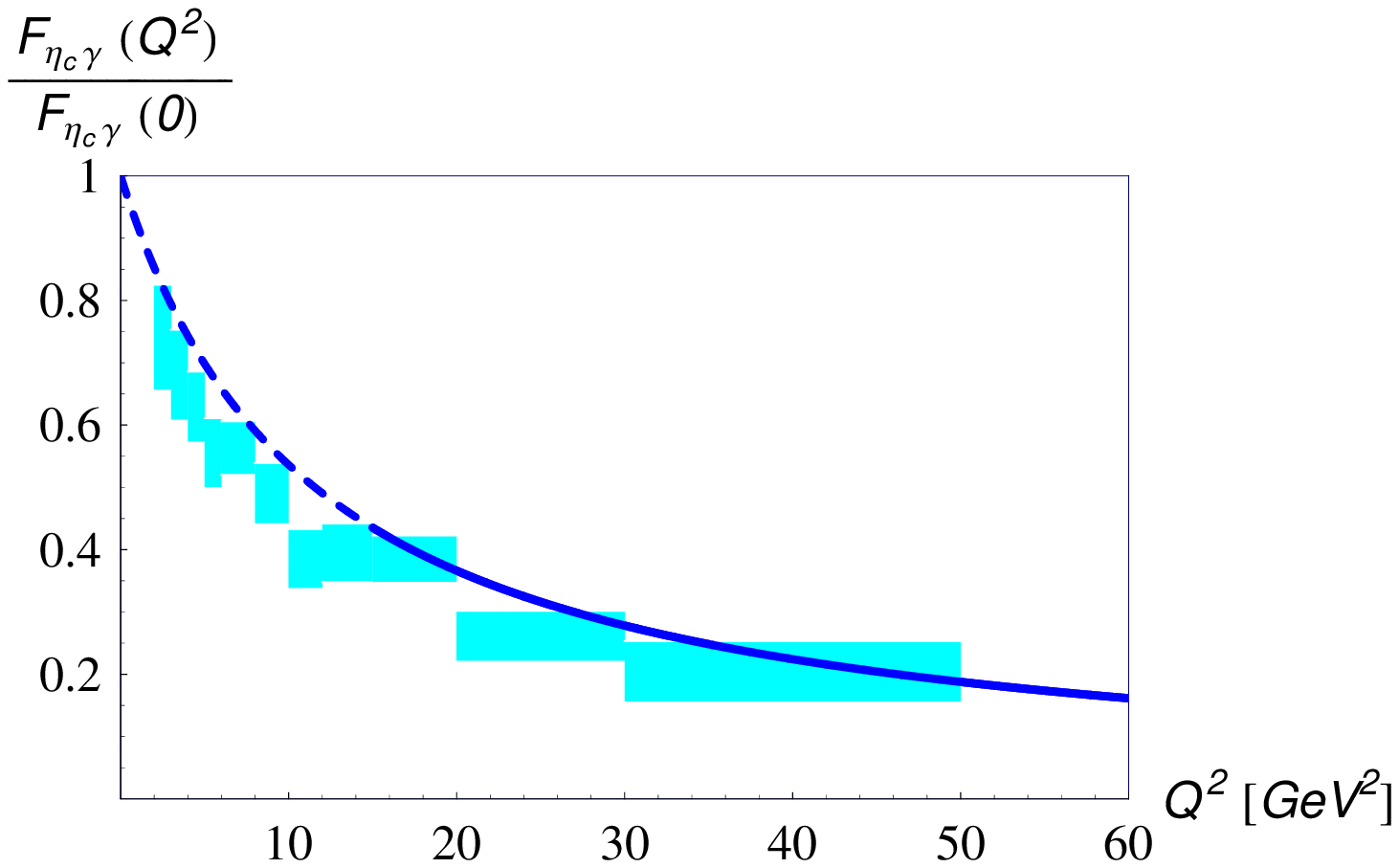}
\caption{Form factor for the transition
$\gamma\,\gamma^*\to\eta_c$: Exact effective thresholds $s_{\rm
eff}^{AVV}(Q^2\to\infty,\beta)$ (top left) and $s_{\rm
eff}^{PVV}(Q^2\to\infty,\beta)$ (top right); form factors obtained
for finite $Q^2$ from the LD sum rules for the correlators
$\langle AVV\rangle$ and $\langle PVV\rangle$ (bottom left); LD
model for the correlator $\langle PVV\rangle$ fitting to {\sc
BaBar} data \cite{babar2010} (bottom~right).}\label{Fig:1}
\end{figure}

For bound states composed of heavy
quarks, the quark mass can no longer be neglected. Finite quark
masses provide an option to exploit not only the correlator
$\langle AVV\rangle$ [as in Eq.~(\ref{AVV})] but also the
correlator $\langle PVV\rangle$ \cite{lm2011}, with, in each case,
an LD model of its own; pQCD factorization then predicts $s_{\rm
eff}(Q^2\to\infty,\beta)$ for both $\langle AVV\rangle$ and
$\langle PVV\rangle.$ Figure \ref{Fig:1} summarizes our findings:
the \emph{exact\/} effective thresholds $s_{\rm
eff}^{AVV}(Q^2\to\infty,\beta)$ and $s_{\rm
eff}^{PVV}(Q^2\to\infty,\beta)$ in the $\langle AVV\rangle$ and
$\langle PVV\rangle$ sum rules differ in their behaviour from each
other as well as from the effective thresholds of relevant
two-point correlators. The LD \emph{assumption\/} $s_{\rm
eff}(Q^2,\beta)=s_{\rm eff}(Q^2\to\infty,\beta)$ entails the
form-factor behaviour depicted in~the bottom row of
Fig.~\ref{Fig:1}. For very small $Q^2,$ this simple LD model
cannot be expected to be applicable. Interestingly, it yields
$F_{\eta_c\gamma}(Q^2=0)=0.067\;\mbox{GeV}^{-1}$ from $\langle
AVV\rangle$ and $F_{\eta_c\gamma}(Q^2=0)=0.086\;\mbox{GeV}^{-1}$
from $\langle PVV\rangle,$ in reasonable agreement with the
measured value
$F_{\eta_c\gamma}(Q^2=0)=0.08\pm0.01\;\mbox{GeV}^{-1}.$
Consequently, we feel entitled to conclude that the LD evaluation
of, at least, the correlator $\langle PVV\rangle$ provides
reliable predictions for a broad $Q^2$ range starting at very low
$Q^2$ values (see also Ref.~\cite{kroll}).

\section{Form factor for the transitions
$\gamma\,\gamma^*\to(\eta,\eta^\prime)$}
\begin{figure}[t]
\includegraphics[scale=.506]{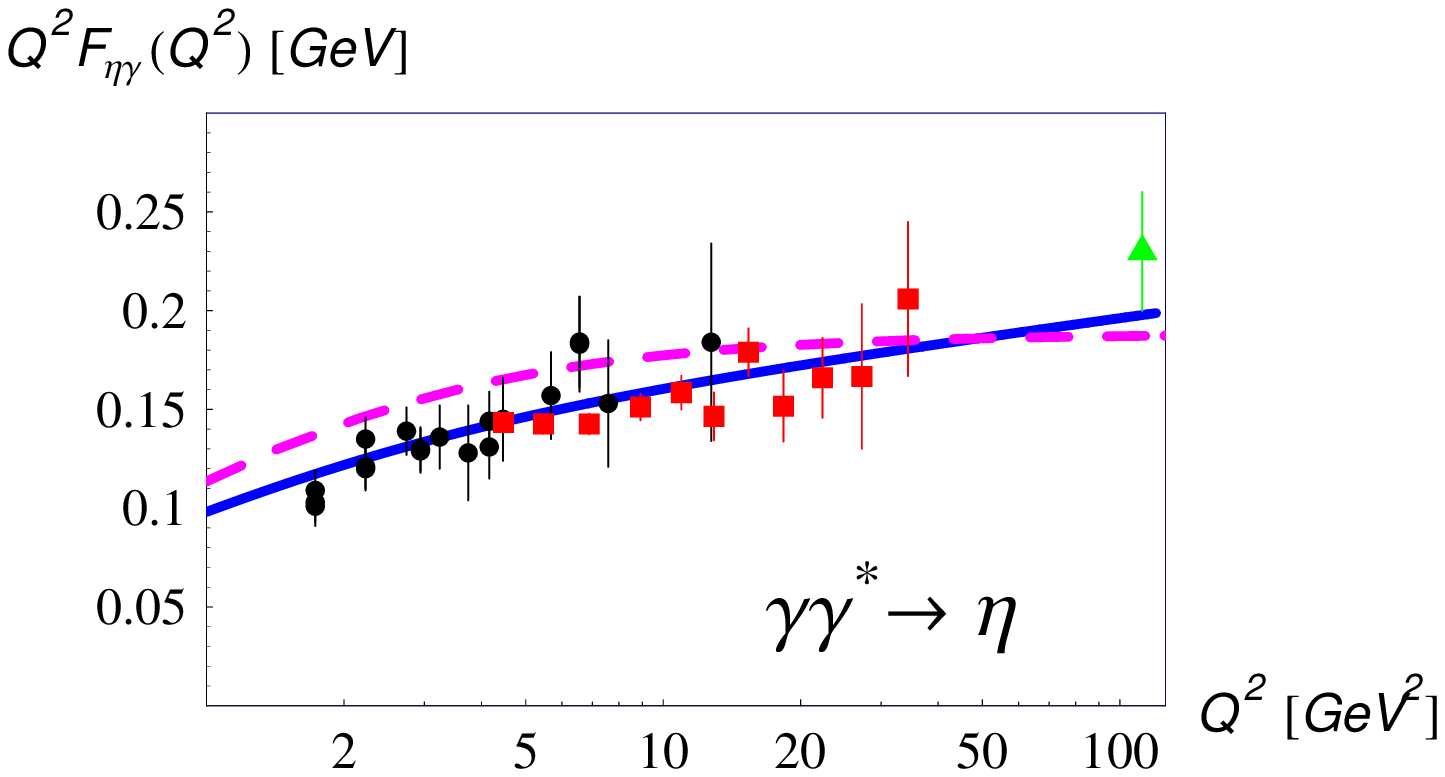}
\includegraphics[scale=.506]{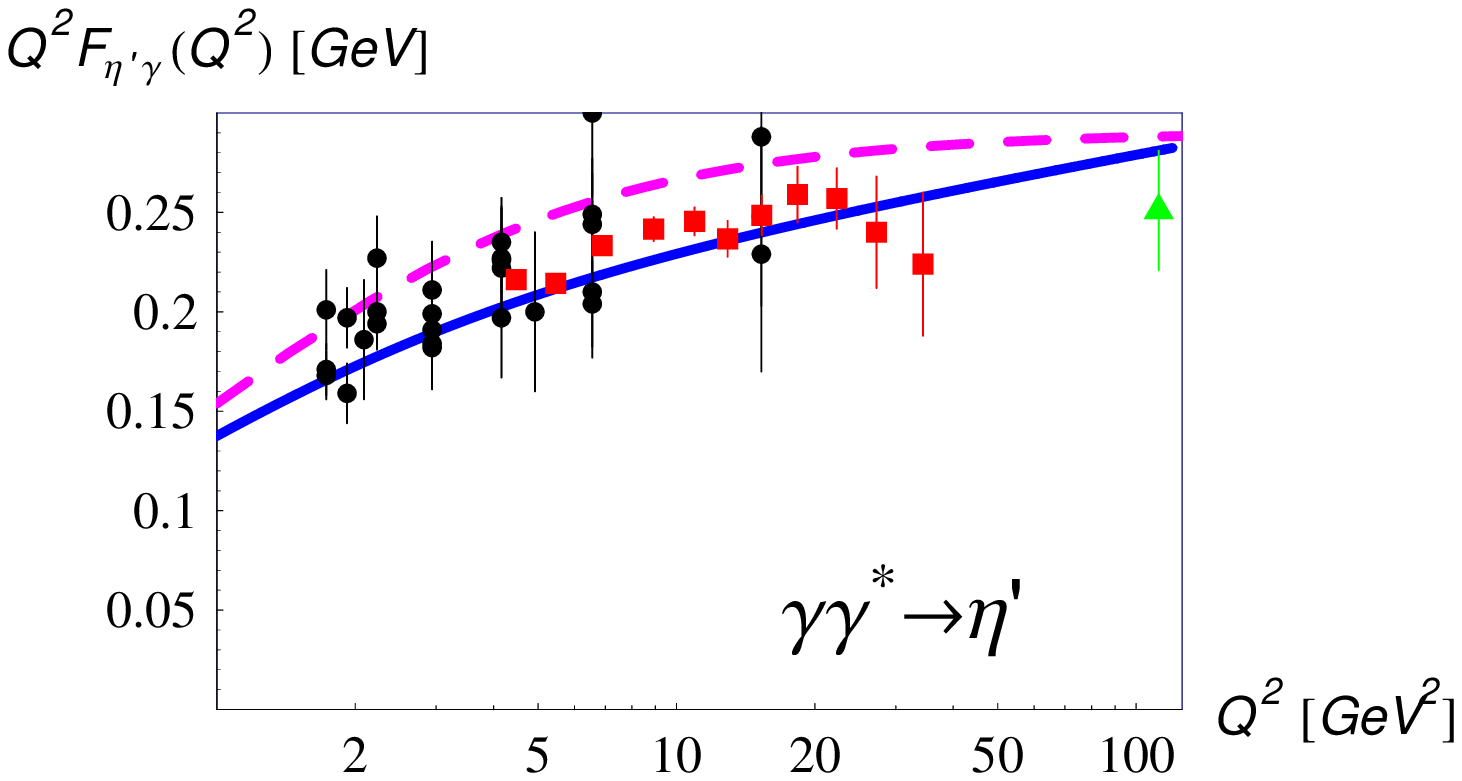}
\caption{Form factors $F_{(\eta,\eta^\prime)\gamma}(Q^2)$ for the
transitions $\gamma\,\gamma^*\to(\eta,\eta')$: LD predictions
\cite{blm2011,lm2011} (dashed lines) and fits \cite{ms2012} (solid
lines) to measurements by CELLO and CLEO \cite{cello-cleo} (black
dots) and {\sc BaBar} (red dots) \cite{babar1}.}\label{Fig:2}
\end{figure}

For $\eta^{(\prime)}$
transitions, the form factors $F_{(\eta,\eta')\gamma}(Q^2)$ are
mixtures \cite{mixing} of nonstrange contributions
$F_{n\gamma}(Q^2),$ with $n$ abbreviating $(\bar uu+\bar
dd)/\sqrt{2},$ and strange contributions $F_{s\gamma}(Q^2),$ with
$s$ indicating~$\bar ss$:
$$F_{\eta\gamma}(Q^2)=F_{n\gamma}(Q^2)\cos\phi-F_{s\gamma}(Q^2)\sin\phi,\qquad
F_{\eta'\gamma}(Q^2)=F_{n\gamma}(Q^2)\sin\phi+F_{s\gamma}(Q^2)\cos\phi,$$with
mixing angle $\phi\approx38^\circ.$ Of course, the sum rules in LD
limit for the latter form factors involve two separate effective
thresholds, $s_{\rm eff}^{(n)}=4\pi^2f_n^2$ and $s_{\rm
eff}^{(s)}=4\pi^2f_s^2,$ with $f_n\approx1.07f_\pi$ and
$f_s\approx1.36f_\pi$:$$F_{n\gamma}(Q^2)=\frac{1}{f_n}\int_0^{s_{\rm
eff}^{(n)}(Q^2)}{\rm d}s\,\Delta_n(s,Q^2),\qquad
F_{s\gamma}(Q^2)=\frac{1}{f_s}\int_0^{s_{\rm eff}^{(s)}(Q^2)}{\rm
d}s\,\Delta_s(s,Q^2).$$Figure \ref{Fig:2} reveals that LD sum
rules \cite{blm2011,lm2011} and experiment
\cite{cello-cleo,babar1} can live with each other pretty well.

\section{Form factor for the transition $\gamma\,\gamma^*\to\pi^0$}
\begin{figure}[b]
\includegraphics[scale=.489]{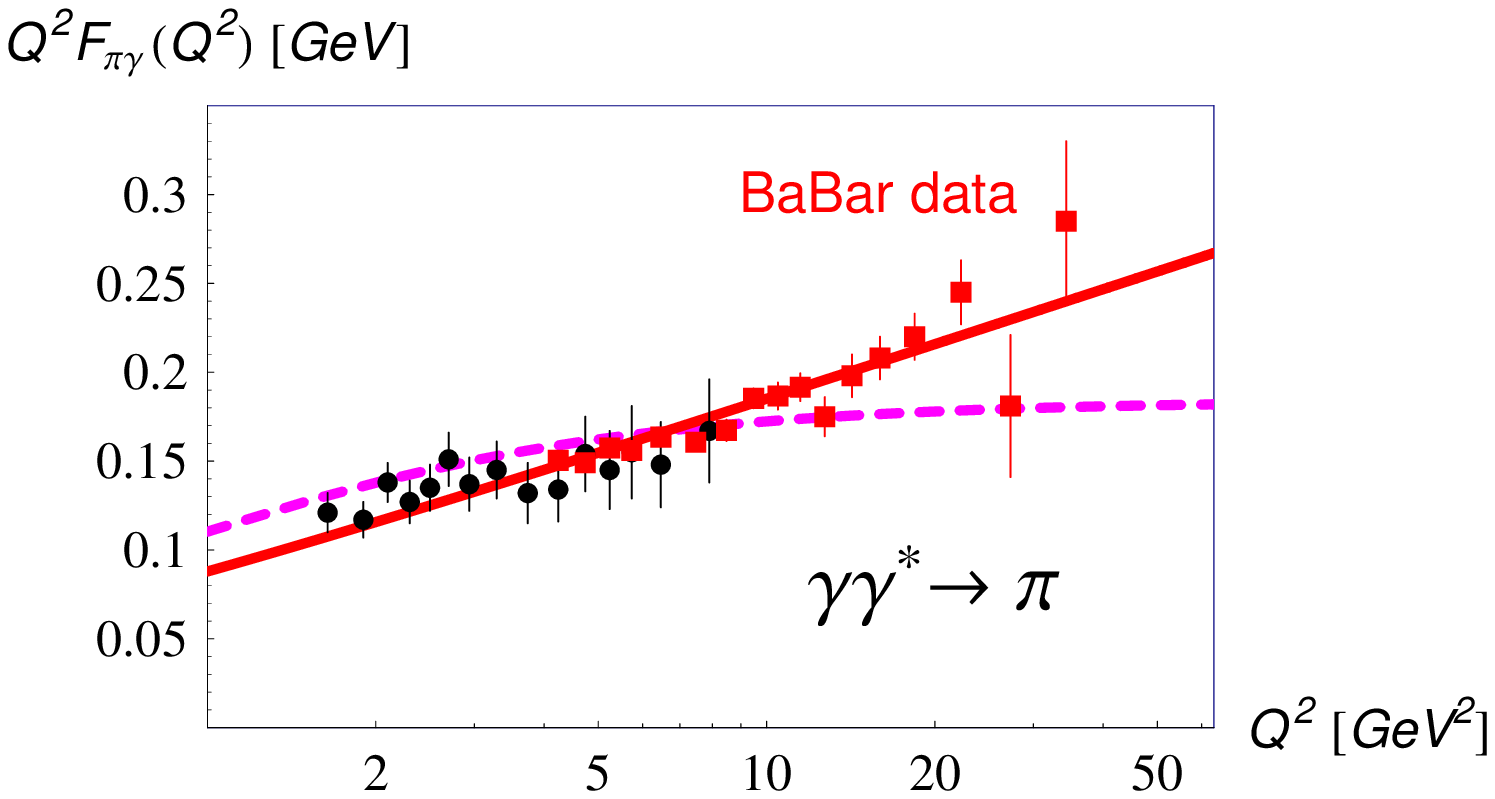}
\includegraphics[scale=.489]{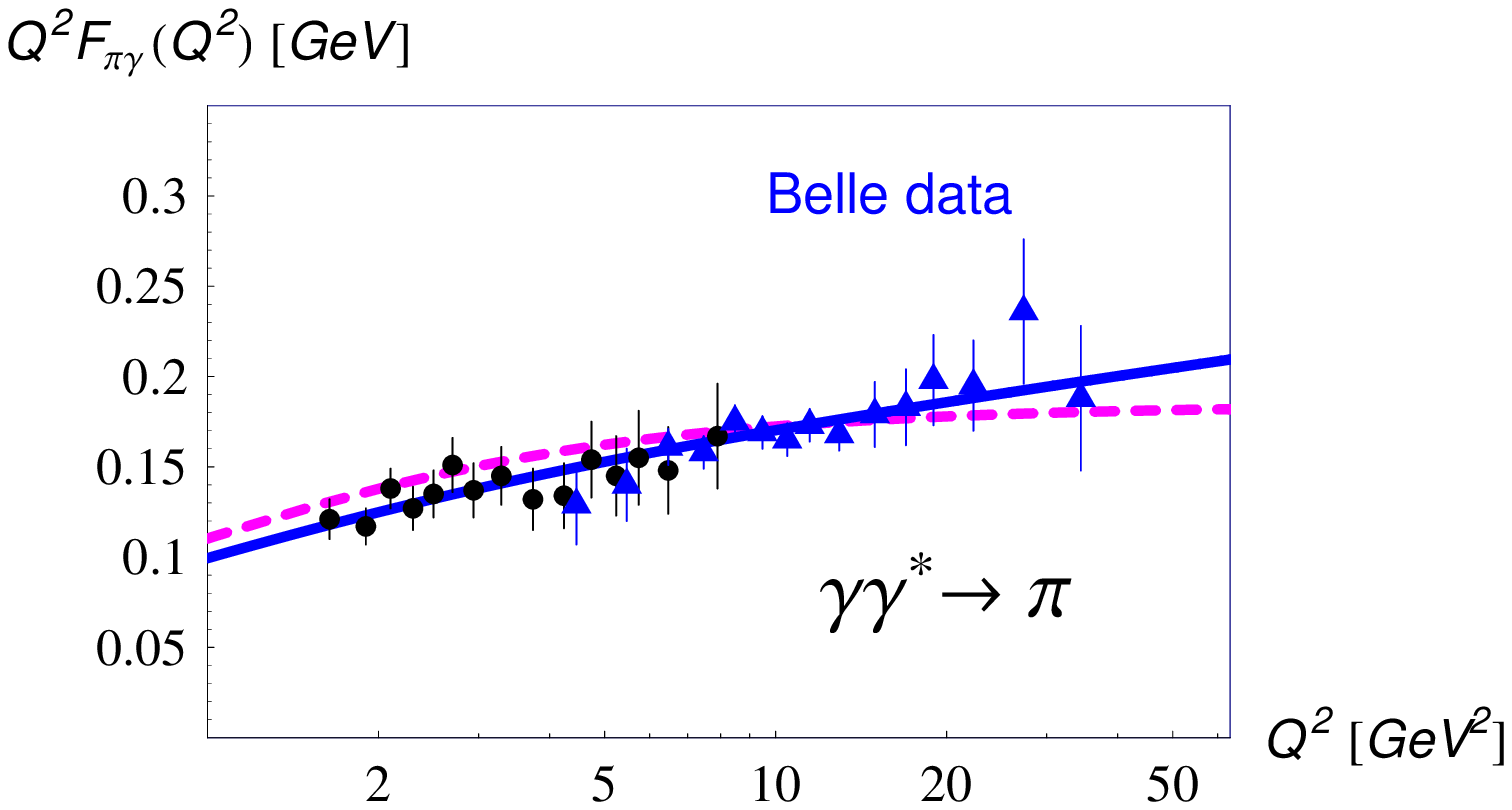}\\
\includegraphics[scale=.489]{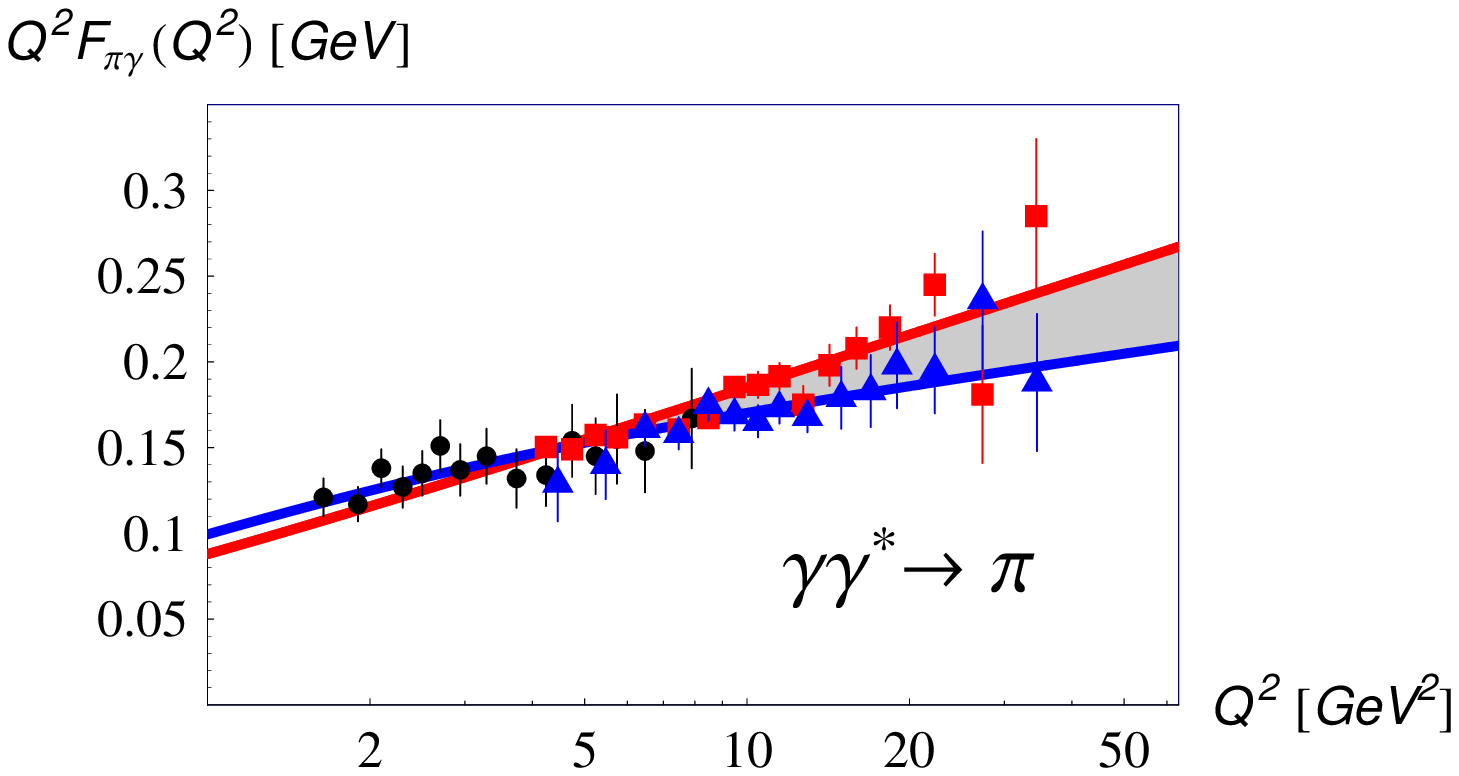}
\includegraphics[scale=.591]{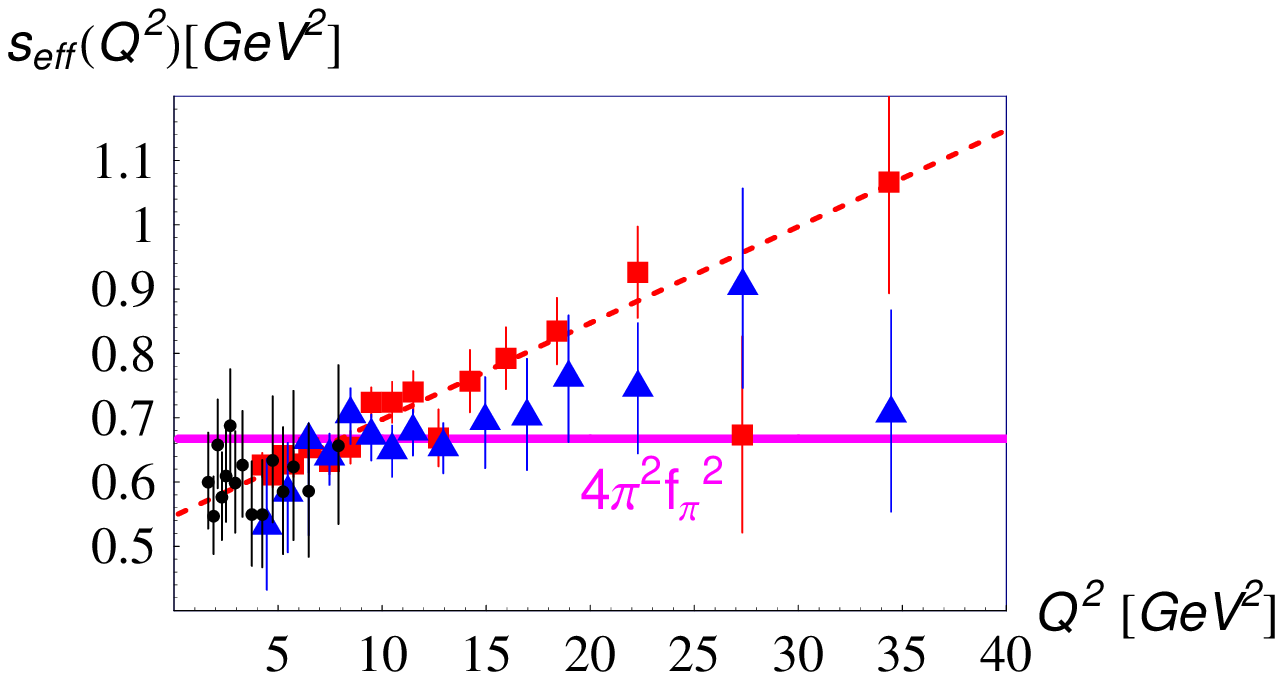}
\caption{Form factor $F_{\pi\gamma}(Q^2)$ for the transition
$\gamma\,\gamma^*\to\pi^0$: CELLO and CLEO \cite{cello-cleo}
(black dots) vs.~{\sc BaBar} \cite{babar} (top left) and Belle
\cite{belle} (top right) data; discrepancy (grey shaded area)
between {\sc BaBar} \cite{babar} and Belle \cite{belle} data for
larger $Q^2$ (bottom left); equivalent effective threshold $s_{\rm
eff}(Q^2)$ inferred for each data point by means of
Eq.~(\protect\ref{srfp}) (bottom right). Magenta lines represent
the LD model, red or blue solid lines a fit
\cite{ms2012}~to~the~data.}\label{Fig:3}\end{figure}

By construction of the sum-rule formalism, the behaviour of any of
the $\pi^0,$ $\eta,$ and $\eta'$ transition form factors in the
limit of large $Q^2$ is governed by spectral densities to be
deduced by evaluating~the relevant pQCD Feynman diagrams;
therefore, it has to be identical for all light pseudoscalar
mesons \cite{ms2012}: The sum rule for the correlator $\langle
AVV\rangle$ in LD limit is equivalent to the anomaly sum rule
\cite{teryaev2}$$F_{\pi\gamma}(Q^2)=\frac{1}{2\sqrt{2}\,\pi^2f_\pi}
\left[1-2\pi\int_{s_{\rm th}}^\infty{\rm d}s\,\Delta^{(I=1)}_{\rm
cont}(s,Q^2)\right],$$with similar relations emerging for the
$I=0$ and $\bar ss$ channels. The behaviour of the spectral
densities $\Delta_{\rm cont}(s,Q^2)$ at large $s$ determines that
of all form factors $F_{\pi\gamma}(Q^2),$ $F_{\eta\gamma}(Q^2),$
and $F_{\eta'\gamma}(Q^2)$ at large $Q^2$ \cite{ms2012}. Now,
quark--hadron duality assumes all $\Delta_{\rm cont}(s,Q^2)$ to be
equal to the associated $\Delta_{\rm pQCD}(s,Q^2)$ in the
respective channel; as purely perturbative quantities, all the
latter must be equal to each other.

Figure~\ref{Fig:3} compares several sets of experimental data
available for the $\pi^0$ transition form factor. The {\sc BaBar}
measurement comes as a great surprise in two respects: On the one
hand, it obviously disagrees with the $\eta$ and $\eta'$ form
factors and with the conventional LD model for $Q^2$ up to
$40\;\mbox{GeV}^2.$ On the other hand, the LD violations claimed
to have been found by {\sc BaBar} rise with $Q^2$ even near
$Q^2\approx40\;\mbox{GeV}^2,$ which is in conflict with hints from
quantum-mechanical analogues. Our confidence in our precursor
studies forces us to conclude that it might be hard to put forward
any interpretation of the {\sc BaBar} results within QCD (see also
the related discussions in Refs.~\cite{findings}). More recent
Belle results for $F_{\pi\gamma}(Q^2),$ although within errors
compatible with their {\sc BaBar} counterparts
(cf.~\cite{agaev,pere}), strengthen our trust: the Belle $\pi^0$
transition form-factor behaviour for large $Q^2$ resembles the
one~of $\eta$ and $\eta'$ and agrees with the expected onset of
the LD regime already in the range $Q^2\ge5$--$10\;\mbox{GeV}^2.$

\section{Conclusions}The form factors parametrizing the amplitudes
for the transitions $P\to\gamma\,\gamma^*$ of the pseudoscalar
mesons $P=\pi^0,\eta,\eta',\eta_c$ have been analyzed within the
framework of local-duality QCD sum rules; there a single key
quantity, the effective continuum threshold, comprises all
nonperturbative effects. Aligning form factors and QCD
factorization yields a threshold model that we regard as
successful:\begin{itemize}\item For all form factors studied,
local duality should perform well for $Q^2$ larger than a few
GeV$^2$:\begin{itemize}\item For the transitions
$\eta\to\gamma\,\gamma^*,$ $\eta'\to\gamma\,\gamma^*,$ and
$\eta_c\to\gamma\,\gamma^*,$ it indeed works reasonably well.\item
For the transition $\pi^0\to\gamma\,\gamma^*,$ {\sc BaBar}
measures a considerable violation of local duality, manifesting by
the effective threshold continuing to rise linearly instead of
approaching \emph{asymptotically} a finite constant, whereas the
trend observed by Belle fits to local duality.\end{itemize}\item
As a whole, the existing experimental data on meson--photon
transitions point towards a tiny residual \emph{logarithmic} rise
of $Q^2F_{P\gamma}(Q^2)$ \cite{ms2012}. If confirmed, this effect
may be interpreted by amending the ratio of hadron-level and
QCD-level spectral densities by an LD-violating~term.\item
Quantum-mechanical experience also leads us to suspect that the LD
sum-rule prediction for the \emph{elastic\/} form factor of the
\emph{charged\/} pion improves with $Q^2$ for
$Q^2\gtrapprox4$--$8\;\mbox{GeV}^2$ and that the corresponding
effective threshold approaches its \emph{asymptotic LD limit\/},
$s_{\rm eff}(Q^2\to\infty)=4\pi^2f_\pi^2,$ already at
$Q^2\approx5$--$6\;\mbox{GeV}^2$ \cite{blm2011}, which is
verifiable by CLAS12 after the JLab $12\;\mbox{GeV}$ upgrade.
\end{itemize}

\vspace{3.23ex}\noindent{\bf Acknowledgments.} D.M.\ is grateful
to B.~Stech for the most pleasant collaboration on the topic of
this talk and to A.~Oganesian and O.~Teryaev for interesting
discussions. D.M.\ was supported~by~the Austrian Science Fund
(FWF) under Project No.~P22843.

\end{document}